\documentclass[10pt,conference]{IEEEtran}\usepackage[]{graphicx}\usepackage[]{xcolor}
\makeatletter
\def\maxwidth{ %
  \ifdim\Gin@nat@width>\linewidth
    \linewidth
  \else
    \Gin@nat@width
  \fi
}
\makeatother

\definecolor{fgcolor}{rgb}{0.345, 0.345, 0.345}

\usepackage{framed}
\makeatletter
 {\par\unskip\endMakeFramed%
 \at@end@of@kframe}
\makeatother

\definecolor{shadecolor}{rgb}{.97, .97, .97}
\definecolor{messagecolor}{rgb}{0, 0, 0}
\definecolor{warningcolor}{rgb}{1, 0, 1}
\definecolor{errorcolor}{rgb}{1, 0, 0}
\newenvironment{knitrout}{}{} 

\usepackage{alltt}

\usepackage{rotating}
\usepackage{statmath}

\usepackage[utf8]{inputenc}

\usepackage{textcomp}

\usepackage{graphicx}
\usepackage{enumerate}

\usepackage{array}
\makeatletter
\renewcommand*{\@fnsymbol}[1]{\ensuremath{\ifcase#1\or *\or \dagger\or \ddagger\or
 \mathsection\or \mathparagraph\or \|\or **\or \dagger\dagger
  \or \ddagger\ddagger \else\@ctrerr\fi}}
\makeatother

\usepackage[american]{babel}
\usepackage{csquotes}
\usepackage[numbers]{natbib}

\newcolumntype{T}{>{\raggedright\arraybackslash}p{1.8cm}}
\newcolumntype{G}{>{\raggedright\arraybackslash}p{3cm}}
\newcolumntype{V}{>{\raggedright\arraybackslash}p{3.7cm}}
\newcolumntype{Z}{>{\raggedright\arraybackslash}p{5cm}}

\usepackage{url}
\usepackage{amsmath}
\IfFileExists{upquote.sty}{\usepackage{upquote}}{}
\begin{document}


\title{Sources of Underproduction in Open Source Software}
\author{Kaylea Champion and Benjamin Mako Hill, University of Washington}
 
 \maketitle

\begin{abstract}
Because open source software relies on individuals who select their own tasks, it is often \textit{underproduced}---a term used by software engineering researchers to describe when a piece of software's relative quality is lower than its relative importance. We examine the social and technical factors associated with underproduction through a comparison of software packaged by the Debian GNU/Linux community. We test a series of hypotheses developed from a reading of prior research in software engineering. Although we find that software age and programming language age offer a partial explanation for variation in underproduction, we were surprised to find that the association between underproduction and package age is weaker at high levels of programming language age. With respect to maintenance efforts, we find that additional resources are not always tied to better outcomes. In particular, having higher numbers of contributors is associated with higher underproduction risk. Also, contrary to our expectations, maintainer turnover and maintenance by a declared team are not associated with lower rates of underproduction. Finally, we find that the people working on bugs in underproduced packages tend to be those who are more central to the community's collaboration network structure, although contributors' betweenness centrality (often associated with brokerage in social networks) is not associated with underproduction.
\end{abstract}


\section{Introduction}

Open source software is frequently supported by teams of developers, system engineers, designers, and support specialists. While these teams are often organized as both firms, they increasingly take the form of networks of self-organized collaborators working together in a model called commons-based peer production \cite{benkler_coases_2002}. 
Although the results of peer production are often innovative and influential (e.g., GNU/Linux, Apache, and Python), the tasks taken on by contributors in these efforts do not always align with tasks that are most needed by the software project's users or by the general public.


Of particular concern is software that is \textit{underproduced}---i.e., low quality, but highly important. Underproduction has been shown to be widespread in open source software \cite{champion_underproduction_2021}.
Some underproduced software may be buried deep in the software supply chain, and vulnerabilities and flaws may not be noticed until they cause disruptions.
How can we identify underproduced software and remediate risks before they cause major failures?

This paper builds on previous software engineering research focused on risk measurement to test a series of hypotheses on correlates and theorized causes of underproduction in open source software oriented to both material conditions (e.g., the programming language used) and social features (e.g., the number of maintainers). We open in §\ref{sec:background} with a review of related work to build intuition around these hypotheses, describe our setting and methods in further detail in §\ref{sec:methods}, and present our analysis in §\ref{sec:results}. We discuss the implications of our results in §\ref{sec:discussion}, describing limitations around these results in §\ref{sec:limitations} before concluding in §\ref{sec:conclusion}. 

\section{Background}
\label{sec:background}
\subsection{The production of Free/Libre Open Source Software}

People start and join free/libre open source software (FLOSS) projects for a wide range of reasons, often including intrinsic motivations \cite{eghbal_working_2020, hannebauer_motivation_2016, lakhani_why_2003}.
The work of these developers can be organized in numerous ways---from individual efforts with few if any other contributors, to casual handovers among whoever is willing to pitch in, to committed small-group collaborations, to networks of thousands of developers coordinating their work and making regular integrated releases \cite{christian_taskbased_2021, crowston_hierarchy_2006}. 
For example, the Apache Project today oversees a widely-used web server but was founded by a group of system administrators who had been informally trading fixes to an older, abandoned piece of software \cite{mockus_two_2002}. The Linux kernel was created by Linus Torvalds as a personal project \cite{benkler_coases_2002} before being shared with the world at no charge. 
The openness and flexibility of how work is organized do not guarantee participation. Volunteers may trickle in slowly, if at all, or be treated so poorly that they leave \cite{miller_did_2022}. Ultimately, software may come to rely heavily on a few individuals or even a single person. Leaders can burn out and may not have a pipeline of candidates to assume key roles \cite{miller_why_2019, tan_scaling_2022}. 

Although FLOSS's impact can be large, the process that creates these goods can be inefficient. Benkler observed that some of the largest and most well-known FLOSS projects function as commons-based peer production organizations that rely on voluntary and self-organized labor \cite{benkler_wealth_2006}. Although a substantial portion of FLOSS is supported by firms \cite{germonprez_rising_2019}, these firms do not generally assign tasks within FLOSS projects. Individuals---especially volunteers---tend to choose their own tasks. Unfortunately, the tasks individuals choose may not be those most needed by users. As a result, underproduction---when the quality of a good falls below its importance---can introduce an important form of risk.

\subsection{Alignment between supply and demand in open source software}

\begin{figure}
\centering
\begin{knitrout}
\definecolor{shadecolor}{rgb}{0.969, 0.969, 0.969}\color{fgcolor}
\includegraphics[width=0.8\linewidth]{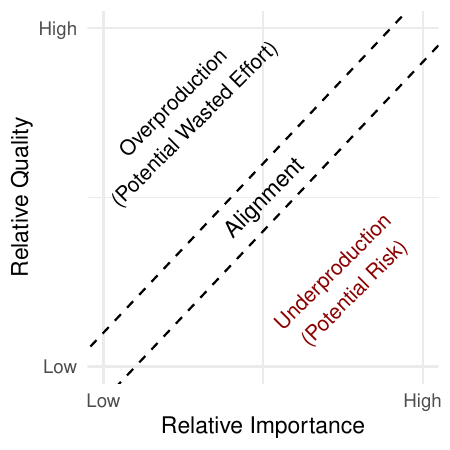} 
\end{knitrout}
\caption{A conceptual diagram locating underproduction in open source software in relation to quality and importance, reproduced from Champion and Hill \cite{champion_underproduction_2021}.}
\label{fig:simAlign}
\end{figure}

In a 2021 paper that has influenced and inspired this work, Champion and Hill proposed a technique for measuring underproduction \cite{champion_underproduction_2021}. Applied to any repository of comparable software packages, this method requires an ordinal measure of quality, an ordinal measure of importance, and a description of an optimal relationship between the two. In the example used by Champion and Hill, the proposed optimal relationship was a simple statement about relative rank: the most important packages ought to be the highest quality (illustrated in Figure \ref{fig:simAlign}). Champion and Hill illustrated their method using data extracted from Debian, finding that a ``minimum of 4,327 packages in Debian are underproduced.'' However, a critical limitation of their work is that it did not examine potential sources, causes, or remedies for underproduction.

\subsection{Sources of Underproduction}

Where does underproduction come from within GNU/Linux distributions? Answering this question is difficult because we lack knowledge in the literature about the role and success of distributions in the software supply chain. However, considering that distributions's work involves high levels of expertise in writing, patching, installing, and integrating software, we can draw lessons from the software engineering literature about what factors could be associated with success and failure. Therefore, to develop hypotheses about the sources of underproduction, we consider both what we know about the most widely cited example of underproduction (the Heartbleed vulnerability in OpenSSL), as well about how underproduction might arise at the software project level, in general.

OpenSSL is open source software and the most widely used implementation of SSL.\footnote{\url{https://heartbleed.com}} The importance of OpenSSL is clear: global web traffic relies on SSL to encrypt traffic between web servers and browsers. In 2014, security researchers identified a vulnerability in OpenSSL that they nicknamed Heartbleed.
At the time of its discovery, the vulnerability had been present for more than 2 years, and security researchers estimated that 24-55\% of the most visited websites were vulnerable \cite{durumeric_matter_2014}. Estimates of the costs to remediate Heartbleed ranged in the hundreds of millions of US dollars \cite{kerner_heartbleed_2014}.
The aftermath of Heartbleed revealed that, despite its importance, OpenSSL had a range of quality challenges \cite{walden_impact_2020, eghbal_roads_2016, Perlroth_heartbleed_2014, pagliery_your_2014}.
The Heartbleed security is emblematic of underproduction because it involves extremely important software that was, at the time, low quality in critical respects.

Scholars have identified several potential reasons for Heartbleed that point to reasons for OpenSSL's underproduction. First, OpenSSL had a complex technical structure that was difficult to analyze by both tools and experts, suggesting a need for widespread refactoring and modernization \cite{walden_impact_2020,wheeler_preventing_2014,wheeler_how_2014,carvalho_heartbleed_2014}. 
Second, OpenSSL is written in C, which, like C++ but unlike Java, lacks built-in detection of buffer over-reads---the bug that led to Heartbleed \cite{wheeler_preventing_2014, wheeler_how_2014}.   
Third, its architecture was substantially out of step with modern engineering standards (e.g., it implemented its own memory management) \cite{wheeler_how_2014}. 
Bug reports related to the dangers of its memory architecture apparently were untriaged in the OpenSSL bug tracker for a substantial period prior to the release of the Heartbleed CVE.\footnote{E.g., \url{https://flak.tedunangst.com/post/analysis-of-openssl-freelist-reuse}}
Finally, there were a relatively small number of people involved with OpenSSL who were available to detect and respond to security vulnerabilities and very little funding devoted to its upkeep \cite{carvalho_heartbleed_2014,eghbal_roads_2016,walden_impact_2020}. 
Although scholars have presented these factors as causes or contributing factors to Heartbleed, the significance of Heartbleed is not simply in the details of how a buffer overread might arise, nor in how such an error might go undetected, but also in the broader problems of neglect and longstanding quality issues in OpenSSL that Heartbleed revealed.

Several of these factors suggest that underproduction is ultimately a technical problem to be prevented and solved. We might expect that modern languages, modern libraries, modern architectures, and modern code analytic techniques could have detected the bug sooner or prevented the bug from being introduced in the first place. Although older packages written in older languages and according to older architectural ideas can be modernized, this is a substantial undertaking \cite{wheeler_preventing_2014,wheeler_how_2014}. 
Examining the problem of underproduction from a technical perspective suggests that some codebases are simply harder to maintain adequately and more failure prone. Technical causes of underproduction might include the language a piece of software is written in, code age, or some combination of these. Although languages and code can be refactored and modernized, the original language and architecture form at least an upper bound on how old the package can be. All things being equal, we might expect newer to be better as we take the opportunity to learn from the past. To test these ideas, we propose three hypotheses: \textbf{(H1) underproduction is associated with older software}, and {\textbf{(H2) underproduction is associated with the use of older programming languages}. Further, given improvements in engineering standards over time and our observation that OpenSSL was both an older package and written in an older language (C), we suggest \textbf{(H3) Underproduction associated with increased package age will be even stronger when the language is also old.}


Although age may be part of the explanation, some software is better maintained than other software of a similar age. There are numerous cases where old software continues to function as well or better than newer software. 
Focusing only on the passage of time would not account for the important role of maintainer resources.
The lack of developers working on OpenSSL is frequently cited as part of the story of how Heartbleed happened \citep[e.g.,][]{walden_impact_2020,eghbal_roads_2016,paul_heartbleeds_2014,pagliery_your_2014}. Proponents of open forms of collaboration point to Linus's law---the notion that ``given enough eyeballs, all bugs are shallow'' \cite{raymond_cathedral_2001}. Presumably, a lack of eyeballs, therefore, leads to undiscovered vulnerabilities. In their study of the PyPi ecosystem, Valiev et al.~found that a higher number of contributors was associated with both short-term and long-term project survival \cite{valiev_ecosystem-level_2018}. A larger OpenSSL maintainer team might have prevented Heartbleed using defect detection tools or seeing the problem sooner.
However, as Brooks famously argued, adding additional developer resources can be detrimental to progress \cite{brooks_mythical_1995}.
In their study of GitHub, Joblin and Apel suggest that small clusters of collaborators with low turnover may be associated with more successful projects in comparison to projects where larger groups of people are committing to the same functions \cite{joblin_how_2022}.
Although we acknowledge the presence of competing explanations, our fourth hypothesis draws from the former perspective emphasizing the value of larger number of contributors and suggests that \textbf{(H4) underproduced software has fewer contributors.}

That said, attributing underproduction to simply the number of developers seems likely to be incomplete, given that FLOSS developers vary widely in such traits as knowledge of the codebase. Turnover in leadership may be harmful, as found in Joblin and Apel \cite{joblin_how_2022}.
In his analysis of a software development firm, Mockus found that staff departures were associated with lower software quality as measured via code analysis and customer defect reports \cite{mockus_organizational_2010}. 
Some open source projects rely heavily on a single individual, and the loss of the maintainer may spell the end of a project. Coelho and Valente surveyed maintainers of GitHub projects that were once popular, but have since been deprecated, and found that lack of maintainer interest and lack of maintainer time were some of the top reasons for project failure \cite{coelho_why_2017}. Further, Coelho and Valente found that although maintainers tried to overcome failure by transitioning ownership to a team, recruiting a new maintainer, or bringing in new contributors, these approaches often did not revive the project.

On the other hand, some FLOSS projects are organized such that high levels of maintainer turnover are not detrimental. For example, in their study of five open source projects composed of many modules in an overarching framework (Angular.JS, Ansible, Jenkins, JQuery, and Rails), Foucault et al.~found that these successful projects all had relatively high contributor turnover \cite{foucault_impact_2015}.
Early work from Michlmayr and Hill observed that Debian tended to rely on single individuals \cite{michlmayr_quality_2003}. Robles et al.~\cite{robles_evolution_2005} found that when a maintainer leaves the Debian project, others often adopt their packages. As a result of this process, Robles et al.~found that the most important packages tend to be maintained by the most experienced maintainers.
However, the adoption of a piece of software by others does not mean that the software will ultimately be maintained at a quality level commensurate with its importance. Nassif and Robillard's 2017 study of eight open source projects (GIMP, Assimp, TrinityCore, Gitlab CE, Linux, Chromium, Kodi, and Apereo CAS) found that the departure of maintainers led to substantial knowledge loss \cite{nassif_revisiting_2017}. Further, Nassif and Robbilard found that these departures led to parts of the project going unmaintained: those who continued to participate did not tend to take on maintenance of the parts written by those who had left.
So, while there are reasons to think otherwise, we hypothesize that the loss of a maintainer will make it more likely that a package will be underproduced and that \textbf{(H5) underproduction is associated with maintainer turnover.}

FLOSS developers organize their work in multiple ways. They may work alone, form a loose collection of collaborators, or assemble into a team \cite{christian_taskbased_2021, crowston_core_2006, crowston_hierarchy_2006, robles_evolution_2005}. Although the team structure does not need to be large or complex to have an impact, the presence of a team suggests that a more stable collaboration has formed with a distinct identity \cite{studer_community_2007}. In their study of the PyPi ecosystem, Valiev et al.~found that having a project hosted in an ``organization'' account, rather than an individual one, was associated with the project being 22\% less likely to become dormant \cite{valiev_ecosystem-level_2018}. Transitioning away from depending on a single maintainer was also one of the techniques Coelho and Valente found that maintainers of failing projects attempted, suggesting that maintainers themselves feel that teamwork may be helpful in preventing failure \cite{coelho_why_2017}. Team approaches offer the potential for the transition of leadership, technical help, and social rewards such as encouragement and recognition. This perspective suggests that \textbf{(H6) packages maintained by a team are less likely to be underproduced than those maintained by individuals.}

Ways of organizing work go beyond the team that forms around a single package. Developers work together across multiple packages and form broad collaborative networks. 
Two software packages inside these networks can be thought of as connected if the same person has engaged with both.
These types of connections may be valuable because someone who works on several packages may be well-placed to see how a bug in one package has implications for another. 
Indeed, Joblin and Apel constructed collaboration networks from developers who commit to the same function and found that degree centrality strongly predicted success for the open source projects they examined on GitHub \cite{joblin_how_2022}.
We might expect that collaboration allows for greater resilience and productivity within the team supporting individual pieces of software because an individual who is highly connected to other collaborators may be able to access additional social and technical support \cite{bird_latent_2008}. This availability of social connection may be highly valuable in driving project success. 
In their study of GitHub projects, Qiu et al.~found that participants were more likely to persist in software projects with high potential for building social capital \cite{qiu_going_2019}. A struggling package that is more central within a collaborative network may be more likely to be noticed, and the resources to solve its problems are more likely to be available. 
These perspectives suggest that \textbf{(H7) underproduced packages are associated with collaborators who are farther from the central influential core of collaborators.} 

Finally, it may be the case that underproduced packages lack collaborators who are aware of what is happening across Debian as a whole---those people who might notice cross-package trends or have a more accurate sense of the baseline rates of problems and resolution such that they can recognize that a package is struggling. This suggests that \textbf{(H8) underproduction is associated with collaborators lacking visibility into what is going on elsewhere in the project---i.e., an absence of brokers.}


\section{Methods}
\label{sec:methods}
\subsection{Empirical Setting} 
We build from Champion and Hill's method for detecting relative underproduction in software packages in the Debian GNU/Linux operating system distribution \cite{champion_underproduction_2021}. 
Debian was founded in 1993 and has grown to serve a range of computing needs, especially with respect to server infrastructure.
Debian also serves as the primary package source for Ubuntu.\footnote{\url{https://ubuntu.com/community/debian}} The Debian community has a history of democratic governance and volunteer participation in that individuals exclusively self-select into tasks and a long-established reputation for quality \cite{oneil_cyberchiefs_2009, sadowski_transition_2008, mateos-garcia_institutions_2008}.

An operating system distribution serves a vital integration role in the world of software development by bringing together, configuring, and testing thousands of packages under a common framework. One of the key ways that Debian members contribute to Debian is by serving as a package's \textit{maintainer}. Although the term is used in several ways within Debian, we use the term to describe the person listed in a package's ``Maintainer'' field. This person (or group) will receive notifications of bugs and is responsible for the package.\footnote{\url{https://wiki.debian.org/Maintainers}} 
In Debian, maintainers are not the only individuals who can update packages. Trusted community members (those granted ``Debian Developer'' status) can upload a Debian package in what is called a non-maintainer upload (NMU). Because they are a sign of problems, Champion and Hill used NMUs to validate their measure of underproduction.
Many packages list a range of other people in package metadata who, while not the maintainer, can upload packages in what are not considered NMUs.
We describe any person who uploads a version of a package outside of an NMU as an ``uploader.''

A Debian package maintainer is not necessarily the person who develops and maintains the software itself. Debian typically describes this latter person as the ``upstream'' maintainer. In this sense, Debian maintainers are the downstream recipients of problems that are located upstream in the software supply chain. However hard a Debian maintainer works, they may have little control over the challenging behavior of the software they are seeking to integrate with other packages. Of course, the responsibilities of the Debian maintainer may be made substantially more difficult if the software they are packaging is complex, has fragile dependencies, or if it not being maintained well (or at all) upstream. Although our study is of Debian, we examine only one supply chain segment. Some of the data we collect directly reflects upstream conditions, which vary widely. Figure \ref{fig:supplyChain} illustrates a piece of this supply chain; upstream software developers produce software, which is packaged by operating system distribution communities like Debian.

\begin{figure*}
    \centering
    \includegraphics[width=.9\linewidth]{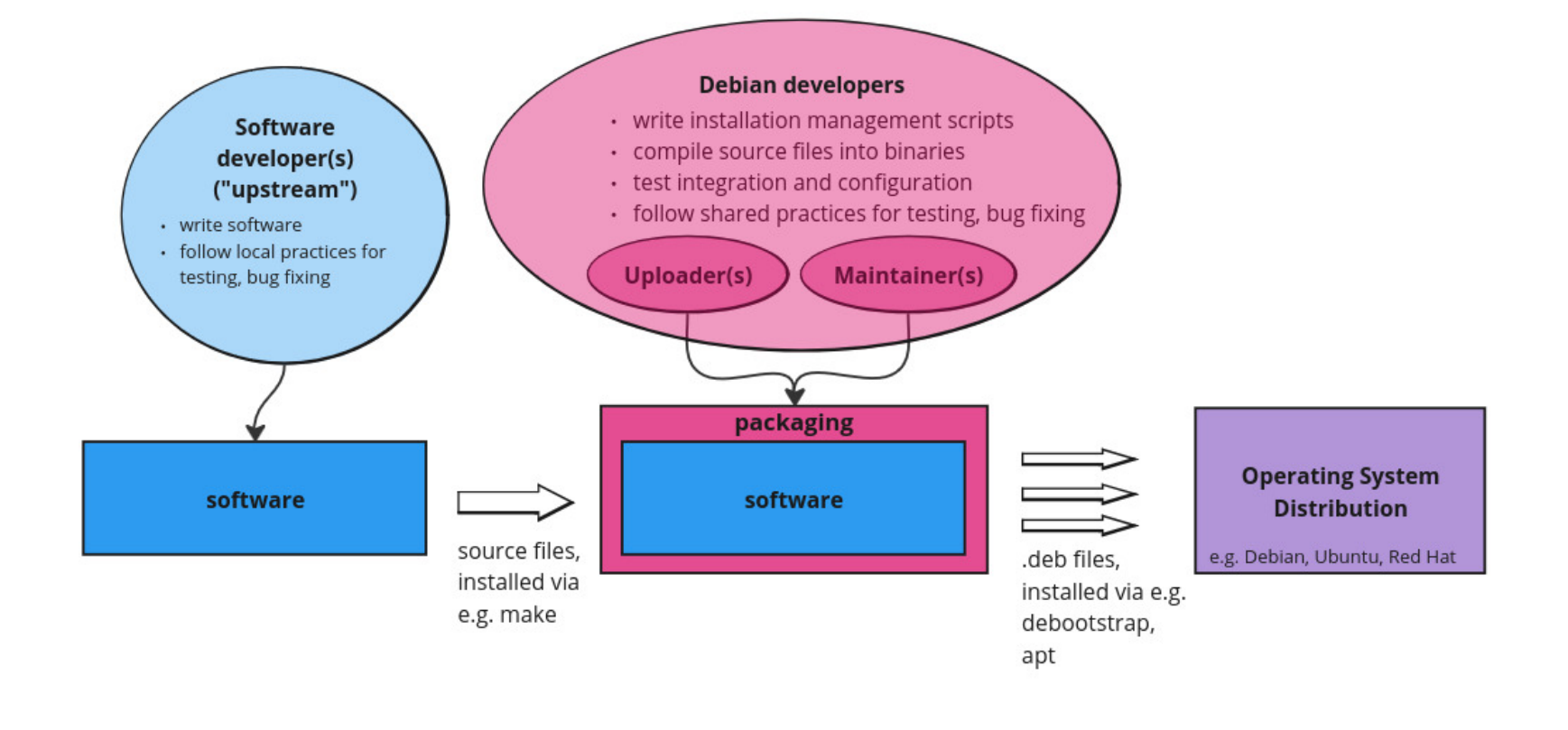}
    \caption{A piece of the free/libre open source software supply chain. Software is typically developed ``upstream'', and then numerous software programs are packaged and integrated by Debian developers before being distributed as part of an operating system or using package management tools. Users may also directly install software from source files or precompiled binaries without the benefit of a package manager (not shown).}
    \label{fig:supplyChain}
\end{figure*}

\subsection{Data}

Debian has a history of making high-quality data available to the public in ways that have been useful for software engineering research \citep[e.g.,][]{caneill_debsources_2016,nussbaum_ultimate_2010}. To test our hypotheses, we operationalize the concepts in each hypothesis using measures from this data (corresponding hypotheses are indicated in parentheses).
We use the history of package uploads as recorded in the Ultimate Debian Database \cite{nussbaum_ultimate_2010} maintained by Debian, as well as package release notes
for package age \textbf{(H1)}, the number of people contributing \textbf{(H4)}, the turnover in maintainership \textbf{(H5)}, and the presence of a maintaining team \textbf{(H6)}. 
We collected package-level underproduction data from the dataset Champion and Hill published.\footnote{ \url{https://doi.org/10.7910/DVN/PUCD2P}} This dataset contains estimates for 6,551 packages \cite{champion_underproduction_2021}.
We merge these data with measures that we build from a series of other datasets. 
We identify the upstream programming language of each package (\textbf{H2}), and we use a set of tags in the UDD of the format ``implemented-in::{\textless}language{\textgreater}''. These tags are present for 2,383 of the 6,551 packages in our study and refer to 24 different languages. 
We also use the history of bug resolutions from the Debian Bug Tracking System (BTS) to build a collaboration network (\textbf{H7}, \textbf{H8}).

\begin{figure}
    \centering
\includegraphics[width=.5\textwidth]{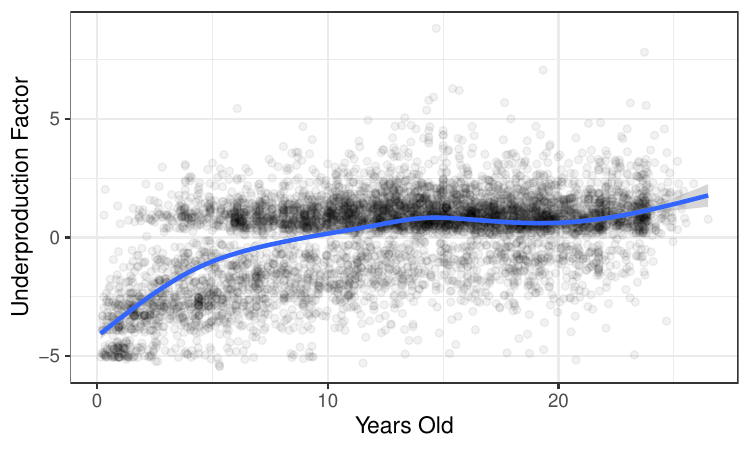}
    \caption{Visualizing package age based on when the package was added to Debian, with a generalized additive model (GAM) line to indicate a moving average.}
    \label{fig:byBirthday}
\end{figure}

\begin{figure}
    \centering
    \includegraphics[width=.5\textwidth]{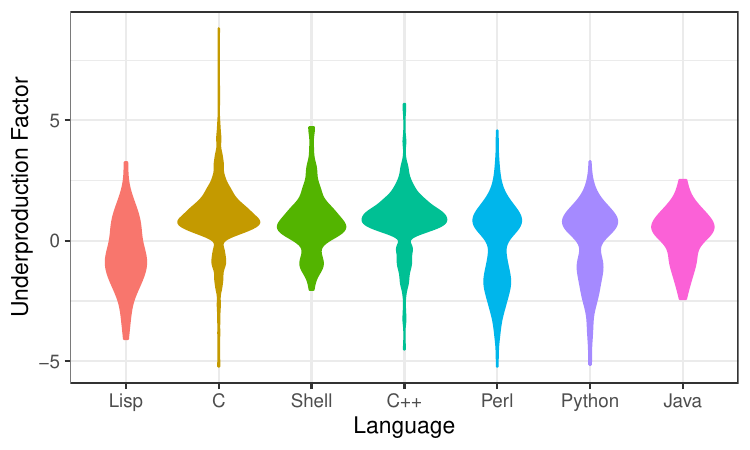}
    \caption{Violin plot of our data distribution broken down by the most commonly appearing languages. See Table I for models which test the relationship between language age and underproduction. This visualization contains data for 2,280 packages. On 135 occasions, the same package appears multiple times because it was consistently tagged as having been implemented in more than one language.}
    \label{fig:byLang}
\end{figure}

\subsection{Measures}

Our key outcome variable, \textit{underproduction}, is drawn from the replication dataset published by Champion and Hill. Positive values indicate underproduction, with a higher underproduction factor associated with higher levels of underproduction; the measure is a ratio between quality rank and importance rank. We treat \textit{underproduction} as a boolean value where true indicates that the underproduction factor, as measured in Champion and Hill, is positive.

In \textbf{H1}, we proposed that underproduction is associated with older software. We operationalize \textit{package age} as the length of time a package has been present in Debian. To do so, we face the challenge of incomplete data because the tracking database was introduced several years into Debian's history. To prevent left-censoring, we take the earlier of two sources of information: the date that first appears in the package changelog file associated with each package (reading the last five lines in reverse to identify the latest date in the file and manually extracting the date in the 160 cases where the parser could not identify a valid date) or the oldest entry in the uploads database. We then subtract the introduction date from our data collection year (2020). As a result, older packages will have a higher \textit{package age}.

For \textbf{H2} about the language age of packages, identifying the language in which a package is written required additional processing because packages are, in some cases, implemented in multiple programming languages and are tagged accordingly. Additionally, packages can change their programming language over time or be tagged incorrectly in ways that are later corrected. Therefore, we calculate \textit{mean language age} per package.
We construct this measure first by measuring language proportion, taking the total proportion of releases tagged with a given language, omitting instances where a release was uploaded without any language tags. In those cases where a single release was uploaded multiple times with differences in tagging, we took the union of the set of languages listed. For example, if a package was uploaded three times, once tagged only as Perl, once tagged Perl and shell, and once tagged only shell, we treated that release as tagged Perl and shell. For a package $i$ and language $j$, 
$$\text{Release Language Proportion}_{i,j} = \frac{\text{TaggedReleases}_{i,j}}{\text{TotalTaggedReleases}_i}$$ 
Hence our measure of language is per-package and per-language, and a continuous value from 0 to 1.

To calculate \textit{mean language age}, we take the number of years before 2020 in which the language was introduced (the year of introduction was sourced from the English Wikipedia article about the language; larger numbers describe older programming languages) and calculate language age at the package level using the proportion of releases tagged with the language (packages can be tagged with multiple languages in a given release), multiplied by the age of the language in 2020. This measure of language age is admittedly very coarse: languages change over time, and new versions may have very different features than were available in older versions. However, we argue that language authors desire to maintain some level of continuity between versions and that the design decisions made early in a language tend to set the overall tenor for what comes later. Because there are 24 different languages represented, for a package $i$ and language $j$: 
$$\text{Mean Lang Age}_i = \frac{\sum_{j=1}^{24}\text{Release Lang Prop}_{i,j} *\text{YearsOld}_j}{\# \text{Total Languages Present}_i} $$

In \textbf{H4}, 
we proposed that underproduced packages will have fewer contributors. We measure the number of people contributing using \textit{uploader count},
the number of people contributing to a package by counting the number of unique individuals who have uploaded new versions of the package. 

In \textbf{H5}, we proposed that underproduced packages will have higher rates of maintainer turnover. 
We measure maintainership turnover by looking at the maintainer field of all contributions uploaded for a given package. It is important to recall that the maintainer of a package may or may not be the person doing the upload. We use a boolean value for turnover so that if a package has had more than one maintainer, \textit{maintainer turnover} is true, otherwise false. 

In \textbf{H6}, we hypothesize that packages maintained by teams will be less likely to be underproduced.
Examples of teams are the Debian Games Team and the Debian Python Team.
In Debian at the time of upload, the Debian archive system logs the identity of the uploader as well as the identity of the maintainer in the form of a name and email address pair. To detect whether a given upload occurred while the project was being maintained by a team, we examine the maintainer field of the log. If the maintainer is a mailing list (indicated by the mailing list email domain lists.alioth.debian.org), we record it as a team. We also identified three addresses representing the Debian-wide quality assurance teams. Because these teams are typically used a placeholder or temporary maintainers, we did not treat the presence of these QA groups in the maintainer field as indicative of being maintained by a team and omitted these uploads from our dataset. 

In \textbf{H7} and \textbf{H8}, we hypothesize that underproduced packages will tend to be worked on by people in less central positions within the broader collaborative network in Debian. 
To build a network that considers both packages and contributors, we considered that contributors may have some social connection by having worked on bugs within the same package.

Using the package \textit{igraph} \cite{csardi_igraph_2006}, we first generated a two-mode network from individuals doing bug work in packages. Two-mode networks are used when two types of nodes exist in a network analysis \cite{wasserman_social_1994}. In our case, one set of nodes represents individuals, and the other set represents packages. The network is formed by generating an undirected edge between people and packages when they work on bugs in the package and an undirected edge between people who have worked on bugs in the same package. 
Our network included every bug in every package in Debian. 
We excluded from this network the internal-only automatic messages (e.g., ``owner@bugs.debian.org'') as these do not indicate actual contributors. 

To obtain network measures at the level of the package, we then project this two-mode network to a single-mode network such that packages are connected to other packages by means of the people who contribute to their bugs. Therefore, packages are ultimately connected by means of having contributors in common. 
This analysis yielded 78,955 contributors across 18,399 packages. 
Not all of these packages were present in the dataset from Champion and Hill due to insufficient bug resolution data to generate an underproduction estimate. Therefore, although we use all 18,399 packages in generating our network measures, our inference is limited to the 6,551 in Champion and Hill.
We use eigenvector centrality (which ranges from 0 to 1) to evaluate \textbf{H7} (proximity to the core). Eigenvector centrality is a weighted form of degree centrality: after assigning weights to nodes based on the number of connections to other nodes, the value of each connection is reweighted such that connections from high-degree nodes are worth more than those from low-degree nodes in a way that is similar to the Google PageRank algorithm. We use betweenness centrality to evaluate \textbf{H8} (cross-project visibility by means of brokerage). Betweenness centrality is the extent to which a given node lies on the shortest path between other nodes.

\subsubsection{Name Canonification}
Measuring the number of unique people contributing to a package in the form of package uploaders in \textbf{H4}, maintainer turnover in \textbf{H5}, as well as building the networks of bug collaborators in \textbf{H7} and \textbf{H8}, required identifying individuals. Doing this necessitated a process of name canonification. Bugs in Debian are represented as email messages, with addresses presented with a name part and an email part. Contributors sending messages with the system do not always use the same name and email address, and their name and address may change over time. A manual review of bug records revealed numerous examples where an individual submitted a bug with their work or personal address and then worked on the bug and resolved it using their Debian-specific address. As a result, email address was not sufficient as a primary key. Unfortunately, a person's name alone is also not a reliable primary key since some names and nicknames are relatively common globally. The risk is not only from conflating two Johns Smith, but also conflating Jack Johnson and Jennifer Juarez who have have both decided to set their name part of their email address to ``JJ.''

Given these concerns, we applied two heuristics as part of our canonification process. First, we reasoned that in all cases, someone using the same email address but a different name was likely to be the same person (e.g., ``J Doe $<$jdoe@example.com$>$'' should be treated as the same person as ``Jane Doe $<$jdoe@example.com$>$''). Further, within the same bug (but only within the same bug), we treated entries with the same name part but different email address part as also the same, e.g., ``Jane Doe $<$jd@workaddress.com$>$" was treated as the same person as "Jane Doe $<$j.doe@homeaddress.com$>$''. 
We also inspected this mapping manually to remove obvious dummy addresses and resolve circular references. This process found 8,741 instances where the same person used more than one address for their Debian bug work and identified aliases associated with 6,438 unique individuals.  
We used these aliases to canonify uploaders, maintainers, and collaborators on bugs prior to analysis.

\subsection{Analytic Plan}
\label{sec:analyticPlan}
We use logistic regression to test each of our hypotheses, which calculates the log odds of an outcome from a vector of explanatory variables. Because the outcome variable in logistic regression is the log odds of an outcome, our models were all of the form: 
$$\log \left( \frac{P_{\text{underprod}}}{1-P_{\text{underprod}}} \right) = \bfbeta_0 + \bfbeta \textbf{X}$$
where \textbf{P}} is the probability that a given package is underproduced, odds is defined as probability divided by 1 - probability, \textbf{X} describes a vector of variables from our dataset, and $\bfbeta$ describes the vector of fitted parameter estimates associated with our hypothesis tests. 

One analytic challenge we faced was missing data in terms of package programming language (due to untagged packages) as well as missing network measures (due to isolates in the collaboration network, meaning we have no defined centrality measure for a package). 
In order to offer inference into the source of underproduction while managing the presence of confounders and missing data, we fit four models. M1 includes those predictors for which we have complete data (offering insight into \textbf{H1}, \textbf{H4}, \textbf{H5}, and \textbf{H6}). M2 omits only the language age predictors (\textbf{H1}, \textbf{H4-H8}). M3 omits only the network predictors (\textbf{H1-H6}). M4 is the full model (\textbf{H1-H8}) but, due to missing data, is estimated using only $\frac{1}{3}$ the observations used in M1.
For each hypothesis, we assess the relationship between our predictors \textbf{X} and underproduction and measure significance at the $\alpha=.05$ significance level. Full code and data for replicating our results are available via the Harvard Dataverse at \url{https://doi.org/10.7910/DVN/N2HIRS}. 

\subsection{Ethics}
This study was conducted entirely using publicly available data published by the Debian community and does not involve any interaction or intervention with human subjects. This type of research using these data has been reviewed by the IRB at the authors' institution and has been determined not to be human subjects research. However, we recognize that this work removes observational data from its original context. Therefore, our publication does not include information that would identify Debian contributors. 

\section{Results}
\label{sec:results}
The results of our models are presented in Table \ref{tab:monster}. 

\begin{table*}
\caption{These logistic regression models assess the extent to which underproduction is a function of a range of social and technical factors. Coefficients are untransformed log-odds estimates with a 95\% confidence interval indicated in brackets. Note that the number of observations varies per model due to missing data.\label{tab:monster}}
 \begin{tabular}{l c c c c} \hline  & M1: no lang/network measures & M2: No language measures & M3: No network measures & M4: Full model \\ \hline (Intercept)                     & $-1.90^{*}$       & $-1.66^{*}$       & $-6.57^{*}$       & $-7.28^{*}$       \\                                 & $ [-2.07; -1.73]$ & $ [-1.91; -1.41]$ & $ [-8.24; -4.89]$ & $ [-9.06; -5.50]$ \\ Package Age (years)             & $0.14^{*}$        & $0.08^{*}$        & $0.32^{*}$        & $0.34^{*}$        \\                                 & $ [ 0.13;  0.15]$ & $ [ 0.06;  0.09]$ & $ [ 0.22;  0.43]$ & $ [ 0.23;  0.45]$ \\ Uploader Count                  & $0.21^{*}$        & $0.13^{*}$        & $0.26^{*}$        & $0.18^{*}$        \\                                 & $ [ 0.17;  0.24]$ & $ [ 0.09;  0.17]$ & $ [ 0.21;  0.32]$ & $ [ 0.12;  0.24]$ \\ Did maintainer change?          & $0.32^{*}$        & $0.35^{*}$        & $0.27^{*}$        & $0.22$            \\                                 & $ [ 0.19;  0.45]$ & $ [ 0.19;  0.51]$ & $ [ 0.03;  0.51]$ & $ [-0.03;  0.47]$ \\ Team proportion                 & $0.17^{*}$        & $0.03$            & $-0.48^{*}$       & $-0.20$           \\                                 & $ [ 0.02;  0.33]$ & $ [-0.17;  0.23]$ & $ [-0.79; -0.16]$ & $ [-0.54;  0.13]$ \\ Eigenvector Centrality          &                   & $16.74^{*}$       &                   & $18.91^{*}$       \\                                 &                   & $ [13.08; 20.41]$ &                   & $ [14.18; 23.64]$ \\ Betweenness Centrality          &                   & $-0.00$           &                   & $-0.00$           \\                                 &                   & $ [-0.00;  0.00]$ &                   & $ [-0.00;  0.00]$ \\ Mean Language Age               &                   &                   & $0.15^{*}$        & $0.16^{*}$        \\                                 &                   &                   & $ [ 0.11;  0.19]$ & $ [ 0.12;  0.20]$ \\ Package Age : Mean Language Age &                   &                   & $-0.01^{*}$       & $-0.01^{*}$       \\                                 &                   &                   & $ [-0.01; -0.00]$ & $ [-0.01; -0.01]$ \\ \hline AIC                             & $6586.11$         & $4373.82$         & $2305.08$         & $2088.71$         \\ BIC                             & $6619.97$         & $4418.64$         & $2345.35$         & $2140.37$         \\ Log Likelihood                  & $-3288.05$        & $-2179.91$        & $-1145.54$        & $-1035.35$        \\ Deviance                        & $6576.11$         & $4359.82$         & $2291.08$         & $2070.71$         \\ Num. obs.                       & $6450$            & $4459$            & $2328$            & $2299$            \\ \hline \multicolumn{5}{l}{\scriptsize{$^*$ Null hypothesis value outside the confidence interval.}} \end{tabular}

\end{table*}

\subsection{H1, H2, and H3: Age}

In \textbf{H1}, we proposed that underproduction would be associated with older software. 
Our results in Table \ref{tab:monster} provide support for this hypothesis. This finding is consistent across all four models. 
We find a significant relationship between the number of years a package has been in Debian and underproduction---the older a piece of software is, the higher its underproduction risk. A visualization of our data with respect to language age appears in Figure \ref{fig:byBirthday}. The coefficients from a model to test this relationship are presented in log-odds units in Table \ref{tab:monster}, and the individual coefficients should be interpreted as the log-odds ratio associated with a one-unit change in the predictor, all other factors being equal. Thus our model M4 predicts that a package older by one year has an odds ratio of 1.406, corresponding to odds of being underproduced that are 40.6\% higher.\footnote{Because our models are logistic regressions, $e^\beta$ is the odds ratio associated with a one-unit change in $\beta$. In this case, $e ^{.34}=1.405$. We do similar transformations for each of the other parameter estimates below.}

In \textbf{H2}, we proposed that packages written in older languages would be more likely to be underproduced. Figure \ref{fig:byLang} shows the data associated with this measure. Our results in Table \ref{tab:monster} provide support for this hypothesis across all models fit with this predictor (M3 and M4). M4 predicts that a package written in a language older by one year, all other factors being equal, has odds of being underproduced that are 17.3\% higher.

\begin{figure}
    \centering
    \includegraphics[width=.4\textwidth]{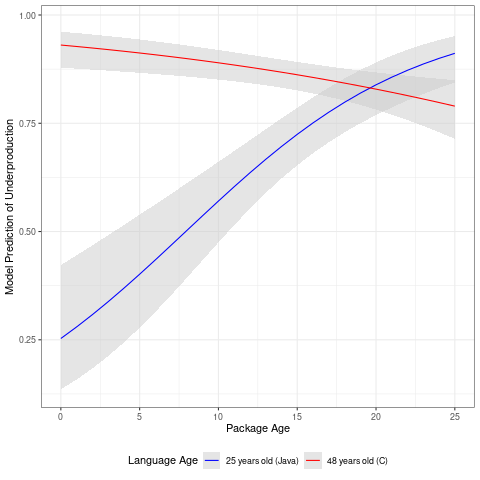}
    \caption{This visualization shows predicted underproduction probability from model M4 for two prototypical packages of different programming language ages where package age varies as shown along the $x$-axis. The package shown in blue is  25 years old, corresponding to a package written in a language as old as Java, while the package shown in red is 48 years old, corresponding to a package written in a language as old as C. The gray ribbon shows a 95\% confidence interval around the prediction.}%
    \label{fig:age_interaction}
\end{figure}

Our hypothesis in \textbf{H3} suggests the presence of an interaction between language age and package age, such that underproduction associated with older packages will be more extreme when the language it is written in is also old. However, our models contradict this hypothesis, and we find that while the main effect of being an older package and being written in an older language both tend to increase the probability that a package is underproduced, the interaction between these two independent variables has a negative coefficient. To interpret these coefficients, we visualize the marginal effect of package age for two language ages (25 years, corresponding to Java, and 48 years, corresponding to C), with results as seen in Figure \ref{fig:age_interaction}. Although the confidence intervals around the predictions are relatively wide, this result suggests that the effect on underproduction of being an older package is weaker when the language is also old.

\subsection{H4: Contributor Count}

In \textbf{H4}, we proposed that greater numbers of contributors would decrease the probability that a package is underproduced. Our results across all four models in Table \ref{tab:monster} contradict this hypothesis. Instead, we find that additional uploaders are associated with an increase in odds that a package is underproduced. Our model M4 predicts that an additional uploader is associated with a 19.7\% higher odds of being underproduced.
In other words, increased uploaders are associated with an increased risk of underproduction. 

\subsection{H5: Maintainer Turnover}
In \textbf{H5}, we proposed that maintainer turnover would be associated with a higher probability of underproduction. Although three of the four model results (M1-M3) in Table \ref{tab:monster} provide support for this hypothesis, we observe that once we include collaboration network measures, language age, and the interaction of package and language age, this effect is no longer statistically significant. This suggests that having experienced maintainer turnover is not associated with higher odds of underproduction once language age and collaboration network measures are held constant.

\subsection{H6: Organizing into Teams}
In \textbf{H6}, we proposed that being maintained by a team would diminish the likelihood that a given package was underproduced. Model results presented in Table \ref{tab:monster} provide little support for this claim. Instead, we find contradictory results in models M1--M3.  With all covariates present in the smaller dataset in M4, we find that the impact of team proportion is not statistically significant. This suggests that, other factors being equal, the involvement of a maintenance team does not impact the odds of a package becoming underproduced. Indeed, our results provide some evidence that the opposite may be true.

\subsection{H7 and H8: Collaboration Networks}
Finally, we examine the role of collaboration networks in underproduction. 
In \textbf{H7}, we test whether underproduced software packages are associated with an absence of influence (eigenvector centrality). Our results for M2 and M4 presented in Table \ref{tab:monster} contradicts our expectation in \textbf{H7}. Instead, we find that an increase in eigenvector centrality is associated with an increased likelihood of underproduction.
In \textbf{H8}, we test whether underproduced software packages are associated with an absence of brokerage (betweenness centrality). Our results for M2 and M4 presented in Table \ref{tab:monster} contradict our expectation in \textbf{H8}. Instead, we find that all other factors being equal, the betweenness centrality of a package is not associated with an increase in the odds that the package is underproduced. 

\section{Discussion}
\label{sec:discussion}

\subsection{The Role of Technology Choices in Underproduction}
In \textbf{H1-H3}, we examined the relationship between older software, older languages, and underproduction. Our results suggest that we should think of software age in a nuanced way. All other things being equal, an older package or one written in an older language is more likely to be underproduced. However, having been written in an older language is associated with a weaker effect of package age. These results suggest that although software faces increased underproduction risk as it (and its language) grows older, thinking of software only in terms of language or age is insufficient. Indeed, for older software in older languages to be present in our dataset, it must have survived for decades. Recently written packages in newer languages are also less likely to be underproduced. Further, we observe that there is substantial variation within languages in Figure \ref{fig:byLang} and in the confidence interval width in Figure \ref{fig:age_interaction}. Committed communities may be able to maintain the health of pieces of software regardless of their age and language, but there are no guarantees.



From the perspective of software users, underproduction risk due to age is likewise challenging. Older software may have a range of benefits not captured in the direct maintenance of the software's quality: knowledge and trust from a history of use, availability of documentation, integration with other tools, embeddedness in a given process, or the existence of migration paths to an alternative. Change to a new service or paradigm has a cost even when the current solution is performing quite poorly. These factors may continue to elevate the importance of the software even after quality has fallen away substantially. 

\subsection{Organizing to Address or Prevent Underproduction}
In \textbf{H4}, we studied the number of contributors by examining uploader quantity. In \textbf{H5}, we examined maintainer turnover. In \textbf{H6}, we examined the question of how uploaders and maintainers are organized. 
Contrary to our expectations, additional uploaders to a package were associated with increased odds of underproduction. Maintainer turnover, in and of itself, is not associated with underproduction. Nor is the presence of a maintenance team in our full model. These are challenging results for communities seeking to organize effort in ways that resist or prevent underproduction. It may be that identifying a level of modularity where a single uploader can sustain effort for the package's life is helpful (hence, the community should orient itself to retaining that uploader's commitment). 
Awareness of the key role of individual effort and the risk involved in a transition from ``one'' to ``more than one'' may allow these projects to think differently about how they approach scale and burnout prevention. 
Or, it may be the case that detection and remediation are more achievable than prevention.

Our results in \textbf{H7} and \textbf{H8} suggest that the people working on bugs in underproduced packages are influential in the network formed by bug commenters. This suggests that, rather than being isolated from other software, underproduced software is drawing from a highly central resource pool---one that is perhaps spread too thinly. 
However, our cross-sectional approach does not allow us to distinguish if the engagement with influential contributors predates the emergence of underproduction or followed after it.
In sum, these findings suggest that the best-case scenario for a piece of software is to be maintained by a dedicated individual who does not work on many other pieces of software.

\subsection{Key Takeaways for Practitioners}

Our study draws both from the distribution level and from upstream development communities, asking whether underproduction at the distribution level is attributable to technical factors such as the age of the package and the language in which the package is written or to how the distribution organizes effort. Although software developers and maintainers in distributions like Debian take on different roles in the supply chain, they have an important relationship. Software developers benefit from the additional testing, dependency management, visibility, and support that distributions provide. To the extent that their goal is to serve end users, making their package easy for distributions to maintain is in their best interest. For their part, distribution maintainers depend on upstream developers to make good quality software that can be easily installed and readily integrated with the other packages. Both groups have a role to play in preventing and addressing underproduction. 

For distributions like Debian, our findings with respect to organizational structure should be particularly helpful. As described, the best-case scenario may be to support dedicated and focused individuals rather than push for simply ``more eyeballs'' or large volumes of new contributors pitching in casually. Although much of the literature on peer production communities emphasizes the power of these casual contributors as part of the long tail, distributions like Debian are an important counterexample.

For developer communities, our finding that underproduction seems to be an inevitable consequence of age and language age suggests that all projects need to confront the march of time. New projects should be careful about using older languages. However, the negative interaction term brings a sign of hope: longstanding projects may continue to pass the test of time.

\section{Limitations}
\label{sec:limitations}

Our measure of underproduction is extracted entirely from prior results in Champion and Hill \cite{champion_underproduction_2021}. This underproduction measure used the mean resolution time of bugs as a measure of quality and usage as a measure of importance. Other measures of quality and importance could lead to different results. Furthermore, our results should be characterized within the context of how underproduction manifests itself at the distribution level. Although Debian is an important part of the software supply chain, it is only one link in the chain. Different links may be characterized by different concerns. Our work may not generalize beyond the Debian context. 

While our evidence suggests that packaging software is well-served by single individuals, some peer production activities are likely to be impossible without a team effort. Understanding the relative modularity of production tasks and the conditions that make it necessary to set aside the advantages of unitary leadership in favor of collaborative effort is a key area for future work.

Our measure of package age does not consider how long a package existed before it was added to Debian and is thus only a lower bound on the age of the software. Our assessment of language age only considers the year in which a given language emerged. Although this is an important part of the context of a given language and the paradigms under which it was designed, languages evolve, and code is rewritten and refreshed. Our use of language tagging in Debian omits variation in how important a given language may be to a package, and these tags are unlikely to be missing at random. 

We have taken up four different perspectives on ways that people collaborate---maintainership, uploading, declaring a team, and working on bugs. However, each of these measures is relatively coarse and does not take the history of the package into account. This omits multiple forms of variation, which may be important, such as how the team functions. Further, our analysis's cross-sectional nature omits the maintenance structure's history. Teams and overworked individuals may adopt packages because they are underproduced. Or what begins as a team effort may fall into disarray with an individual left picking up the pieces.
We have sought to limit the impact of confounders like these by including a full model with all predictors. However, because our collection of predictors is necessarily incomplete, our ability to infer the causes of underproduction is limited. 

Additionally, we acknowledge that many of our predictors of underproduction are largely measures of quality---either of software or of software maintenance. We hope that by treating underproduction as our outcome, instead of quality directly, we are able to incorporate knowledge about importance to identify the causes and correlates of software that is less high quality \textit{than it should be given its importance}. Doing so means that our analysis is substantively about risk, not only about identifying low-quality software.

A final limitation is the cross-sectional nature of our data and the correlational nature of our analysis. Although our hypotheses are framed in terms of causes, our results describe correlations in our data. Indeed, it is easy to imagine how underproduction could cause increases in some of our measures. For example, a package might have more maintainers over its life because it requires lots of work to fix bugs in a way that burns out maintainers or because the software is so important that it attracts lots of very capable people interested in helping. We have attempted to include a range of covariates to address this risk and have attempted to avoid causal language in our interpretation. That said, our results are best thought of as correlational evidence in support of causal theories.

Future work should seek to further unpack the factors that affect quality or importance to understand how these factors ultimately affect underproduction. Further, we should explore what social and structural factors might affect communities' underlying ability to ensure more or less alignment between quality and importance. Additional methods need to be developed to understand underproduction not only in a cross-sectional and cumulative manner, as in Champion and Hill \cite{champion_underproduction_2021} and our own work, but also longitudinally, to support prioritization and intervention.

\section{Conclusion}
\label{sec:conclusion}


Underproduction is partly a result of the simple passage of time. Older software, or software written in older programming languages, is at greater risk. This makes confronting underproduction risk a seemingly inevitable task for software in the longer term. One strategy to confront underproduction risk is to consider how best to organize maintenance effort. Although solitary contributors and teams may be viable, our results suggest that underproduction risk is associated with projects with higher resources. We find no evidence that maintainer turnover is associated with higher risk or that teams are associated with lower risk in our full model. 
The work of sustaining FLOSS is both an opportunity for individuals to make important personal contributions and for them to band together to build teams within larger communities. Although communities producing FLOSS distributions have little control over the age or language of the software they package, they have control over other things.
Our work points to ways that FLOSS organizations can use information on relative underproduction and its correlates to allocate resources and make difficult choices about when to retire or omit software when further intervention may not address the likely underlying causes of problems.

\section*{Acknowledgments}
The work would not have been possible without the generosity of the Debian community. We are indebted to these volunteers who, in addition to producing Free/Libre Open Source Software software, have also made their records available to the public. We also gratefully acknowledge support from the Sloan Foundation through the Ford/Sloan Digital Infrastructure Initiative, Sloan Award 2018-11356 as well as the National Science Foundation (Grant IIS-2045055). This work was conducted using the Hyak supercomputer at the University of Washington as well as research computing resources at Northwestern University.

\bibliographystyle{IEEEtran}
\bibliography{refs}
\end{document}